\documentclass[11pt]{article}

\usepackage{amsmath}
\usepackage{amssymb}
\usepackage{caption}
\usepackage{graphicx}
\usepackage{authblk}
\usepackage[hypertexnames=false,colorlinks=true,linkcolor=blue,citecolor=blue]{hyperref}
\usepackage[numbers,comma,square,sort&compress]{natbib}
\usepackage[a4paper,text={6.5in,10in},centering]{geometry}

\usepackage[dvipsnames]{xcolor}
\usepackage{soul}
\usepackage{subcaption}

\graphicspath{{eps/}{pdf/}{png/}{pics/}}

\makeatletter\@addtoreset{equation}{section}\makeatother

 {\begin{trivlist} \item[]{\bf Proof. }}%
 {\hspace*{\fill}$\rule{.4\baselineskip}{.4\baselineskip}$\end{trivlist}}

 {\begin{trivlist}\item[]\textbf{Acknowledgments.}}{\end{trivlist}}
\newcommand{\bX}{\mathbf{X}}
\newcommand{\bY}{\mathbf{Y}}

\newcommand{\bq}{\mathbf{q}}

\begin{document}


\title{Identification of an influence network using ensemble-based filtering for Hawkes processes driven by count data}

\author[1]{N.~Santitissadeekorn}
\author[1]{S.~Delahaies}
\author[1,2]{D. J. B.~Lloyd}
\affil[1]{\small Department of Mathematics, University of Surrey, Guildford, GU2 7XH, UK}
\affil[2]{\small Center of Criminology, University of Surrey, Guildford, GU2 7XH, UK}


\maketitle

%


\begin{abstract}

Many networks have event-driven dynamics (such as communication, social media and criminal networks), where the mean rate of the events occurring at a node in the network changes according to the occurrence of other events in the network. In particular, events associated with a node of the network could increase the rate of events at other nodes, depending on their influence relationship. Thus, it is of interest to use temporal data to uncover the directional, time-dependent, influence structure of a given network while also quantifying uncertainty even when knowledge of a physical network is lacking. Typically, methods for inferring the influence structure in networks require knowledge of a physical network or are only able to infer small network structures. In this paper, we model event-driven dynamics on a network by a multidimensional Hawkes process. We then develop a novel ensemble-based filtering approach for a time-series of count data (i.e., data that provides the number of events per unit time for each node in the network) that not only tracks the influence network structure over time but also approximates the uncertainty via ensemble spread. The method overcomes several deficiencies in existing methods such as existing methods for inferring multidimensional Hawkes processes are too slow to be practical for any network over $\sim50$ nodes, can only deal with timestamp data (i.e. data on just when events occur not the number of events at each node), and that we do not need a physical network to start with. Our method is massively parallelizable, allowing for its use to infer the influence structure of large networks ($\sim10,000$ nodes). We demonstrate our method for large networks using both synthetic and real-world email communication data.

\end{abstract}

%

\section{Introduction}

Identifying and modelling a structure of influence in a network is an important task in network analysis. An interpretation of ``influence" may depend on the observed data or types of networks. A common notion of influence assigns nodes who effectively spread information as the influencers. Measuring information spread requires more than knowledge of network connections. For example, in a communication network, if A talks to B, will this action increase more probability that B will talk to other people in the network?
\par
In practice, the content of the communication may have to be considered when quantifying the influence. However, this may be unavailable in most data (e.g. cell phone or email data may contain only timestamps and sender and receiver). Our discussion will focus only on contentless data. A common approach to identifying and modelling influence is the idea of influence maximization based on information cascade\cite{Domingos01,Kempe03}. In particular, it finds a subset of $K$ nodes that maximizes the expected number of eventually reached nodes by information spreading from the selected nodes. This approach is a NP-hard problem, where current research focuses on the scalability of the algorithm for a large-scale network\cite{KO20181,Estevez2007,Chen2010}. Recent advances also incorporate the time taken to spread information from one to another\cite{Chen2021} or deal with a special type of network such as random boolean network\cite{Parmer22}. One of the main deficiencies of this approach is that a physical network must be given as an input to their algorithms while our work, which will be discussed below, requires only a time-series of count data to construct a network with edges weights indicating influence strength. Another well-known approach to studying influence structure in a network is a minimum-spanning temporal tree\cite{Huang2015,Morse2019}. This approach can be used to find a persistent cascade pattern for a temporal network. The temporal network here is constructed from observed data that includes ``sender", ``receiver", timestamps and duration of the event, which is a common format for a variety of data such as cell phone communication. The cascade pattern gives important information to infer the influence structures. This approach incorporates temporal knowledge of information spreading (e.g. ``burst" events are frequently driven by influencers' activity and has a strong effect on information spread\cite{Miritello2011}). It assumes a similar concept to epidemic spreading and aims to find patterns of persistent cascade in information diffusion.
\par
A different methodology based on Hawkes process~\cite{Hawkes74} has also been developed to answer similar questions about identifying and modelling influence network structure. However, its main focus is on constructing an influence network from a time-series of some observable events for individuals. For example, the data may contain only timestamps of emails sent out by A with no knowledge of the recipients. However, if A has a strong influence on B, we may observe unusually busy activities created by B immediately after A's activity. The (multivariate) Hawkes model is widely adopted to model this notion of influence and used in many application contexts (e.g. opinion networks in social science\cite{DeVGBG16}, earthquake networks~\cite{Zhuang04}, email network~\cite{Zipkin16}, or healthcare network~\cite{Zhao2014MiningMR}).
This approach infers a network influence structure from certain parameters of the Hawkes model without the need for knowledge of a physical network; hence, the problem becomes a parameter estimation problem. Note that the number of parameters of the multi-dimensional Hawkes process is typically $O(N^2)$, where $N$ is the number of nodes in the network. This poses a significant challenge for a large-scale network inference. The maximum likelihood principle is normally used, and various algorithms based on expectation maximization (EM) have been developed for this problem\cite{Mohler11,Foxetal16,Zipkin16,Yuan19}. In~\cite{Linderman14}, a fully Bayesian approach has been developed to estimate parameters via Markov Chain Monte Carlo (MCMC) for a non-parametric Hawkes process with a focus on a latent network reconstruction, This approach allows not only to include a prior model for the parameters but also the network structure. The prior network structure can be useful when a notion of ``distance" between nodes is possible (e.g. when nodes represent spatial locations and long-range influence is not expected). A similar Bayesian framework where the posterior mode is estimated by a stochastic EM is developed in~\cite{Iwata13}.  The utility of these existing methods is limited to the timestamp data (i.e. the occurrence time of each event) and may not be feasible for a large-scale network in general. 
\par
\begin{figure}[htbp]
            \centerline{\includegraphics[scale=0.33]{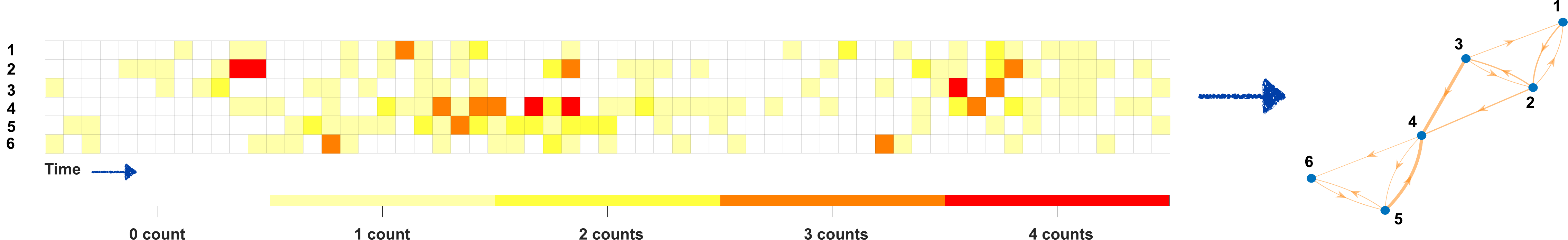}} \caption{Illustration of the influence network reconstruction from the sequence of count data. In this case, there are 5 data sources and the count data is collected over the uniform time interval.}\label{fig:count2nw}
        \end{figure}

In some situations, the timestamp data may be unavailable and only the number of events within a given time interval is aggregated as count data per unit time. The primary goal of this work will introduce a network inference approach for the multidimensional Hawkes process driven by count data, which could encompass an arbitrary number of events in a time interval, is massively parallelisable, able to quantify uncertainty, and does not require knowledge of a physical network. Figure~\ref{fig:count2nw} illustrates the main idea of this work, where sequences of count data from 6 data sources observed in a fixed time interval are used to construct a directed network whose links represent the influence among the nodes. Uncovering directional influence can provide insightful information on how the dynamic of a network might be evolving when activity is created by key influencers. For example, in a so-called covert network (e.g. communication among people involved in criminal/terrorist activities), influential people tend to avoid creating too many communications/activities due to the concern to keep their identities secret. Thus, while the number of activities directly linked to the influential people could be rare, the presence of these activities will be quickly responded to by other nodes, resulting in a cascade of events immediately after the influential activity, which will decay over time.

\par
This work adopts a multidimensional Hawkes process driven by count data, which allows for interactions of two or more one-dimensional Hawkes processes, to describe an influence structure of a network. Each node in the network generates a finite number of events with a conditional rate that changes according to the occurrences of events within the entire network. In particular, the weighted edges of the directed network represent directional influences among nodes that are parameterized by the multidimensional Hawkes model. Therefore, the network inference herein becomes a parameter estimation problem of the multidimensional Hawkes process. Note that the number of parameters of the multidimensional Hawkes process is, typically $O(N^2)$, where $N$ is the number of nodes in the network. This poses a significant challenge for  large-scale network inference.
\par
A reconstruction method for a large-scale influence network would require a trade-off between optimality and computational efficiency. We adopt a recent ensemble approach in~\cite{Santitissadeekorn18} for a sequential data assimilation algorithm for count data. In contrast to most existing methods, which are based on MLE and use a large batch of timestamp data for an inference, this ensemble-based filtering is an online tracking tool that avoids a computationally intensive evaluation of the likelihood function of the large batch data. Thus, it has potential to improve computational scalability for a large-scale problem. In addition, model parameter can often change when applied to real-world data, and a filtering system will allow for such changes to be tracked~\cite{Santitissadeekorn18,Santitissadeekorn20}. Through the ensemble spread, the proposed method will also provide an approximation of uncertainty of conditional intensities and network structure, which could be useful for a decision-making process.

The paper is outlined as follows. In section~\ref{s:dis_hawk}, we describe the multidimensional Hawkes process model for influence and in section~\ref{s:filter} we describe the ensemble filtering to sequentially infer the parameters in the multidimensional Hawkes process and the influence network structure. In section~\ref{s:results}, we present the application of the ensemble filtering method to synthetic networks, including the imperfect model case. We then apply the ensemble filtering method to real-world data in section~\ref{s:realworlddata} and finally draw some conclusions in section~\ref{s:con}.

\section{Hawkes process driven by count data and networks}\label{s:dis_hawk}
This section describes a multidimensional Hawkes process driven by count data, which is in a way a discrete-time version of the traditional continuous time Hawkes process. However, we emphasize that the selection of this model is mostly motivated by its count data driven dynamic, which resonates with the main goal of this work.
\par
We consider a network with $m$ nodes and let $\{\Delta N^j_k\}$ for $j=1,\ldots,m$ and $k=0,1,\ldots$ be the total number of occurrences generated by the $j$-th node during the $k-$th time interval of a length $\delta t$, i.e, $[(k-1)\delta t,k\delta t)$. The history of the number of events up to the $k-$th time interval is denoted by $\mathcal{H}_k:=\{\Delta N^j_1,\ldots,\Delta N^j_k\}$.
We assume that the probability of the number of events within a time interval $\delta t$ follows the Poisson distribution with mean $\lambda^i_{k}\delta t$, where $\lambda^i_k$ denotes a (constant) conditional intensity of the $i-$th node $\lambda^i(t)$ with $t\in[(k-1)\delta t,k\delta t)$. We model the conditional intensity by a doubly stochastic process~\cite{CoxLewisbook} with the following conditional intensity,
\begin{equation}\label{eq:dh}
 \lambda^i_{k+1}=\mu^i+(\lambda^i_k-\mu^i)(1-\beta^i\delta t) +\sum_{j=1}^m\alpha^{ij}\Delta N^j_k,
\end{equation}
where we assume $\lambda^i_{0}=\mu^i$ for all the nodes and $\mu^i,\beta^i,\alpha^{ij}>0$. The parameter $\mu_i$ represents the baseline intensity (which is the lower bound of the intensity process) and $\beta_i$ is the decay rate. The $\alpha^{ij}$ quantify the influence strength that node $j$ has on node $i$. Therefore, the value of $\alpha^{ij}$ is the weight of the edges in the network and represents the strength of the influence that the $j-$th node has on the $i-$th node. In other words, a large value of $\alpha^{ij}$ implies that the $i$-th node has a higher tendency to generate events immediately after one or more generations of events by the $j$-th node. Note that this model is a discrete version of the multidimensional Hawkes process with the exponential decay kernel~\cite{Hawkes74}.

\section{Ensemble filter and uncertainty of the network}\label{s:filter}
 Our primary focus is a sequential data assimilation method in which the number of events is sequentially used to jointly estimate the conditional intensity and the model parameters. The model~\eqref{eq:dh} will play the role of the forecast model for the data assimilation.
 A general-purpose sequential data assimilation algorithm such as particle filter (PF)~\cite{Doucet11tutorial} can be intractable for a high-dimensional problem that one would typically encounter when trying to carry out inference of a large network. Alternatively, the ensemble Kalman filter (EnKF) has been developed to give a suboptimal filtering algorithm and has been widely used in many applications~\cite{EnKFBook06}. Nonetheless, its application for the Poisson count data is not straightforward due to the inherent heteroscedastic variance of Poisson likelihood, e.g., see discussion and numerical examples in~\cite{Santitissadeekorn18}. We adopt the ensemble Poisson-Gamma filter (EnPGF) which has been recently developed in~\cite{Santitissadeekorn18}. The idea of EnPGF is concisely reviewed below, a detailed derivation of some key formula are stated in the appendix.
\par
In a nutshell, the EnPGF algorithm has two main stages. For the first stage, the prior ensemble members of the conditional intensity are moved according to a stochastic equation~\eqref{eq:EnPGFmain} below. The redistribution of the ensemble~\eqref{eq:EnPGFmain} is consistent with the Poisson-gamma conjugacy. The updated ensemble of conditional intensity will then be used as the ``observation" for the subsequent stage. For the second stage, the innovations in terms of the conditional intensity are linearly regressed on the ensemble of corresponding parameters. This algorithm showed good results for several models of the conditional intensity process, see~\cite{Santitissadeekorn18}. The mathematical description of EnPGF and its application to the multidimensional Hawkes process are given below.
\par
To avoid cluttering the notations, we temporarily suspend both time and location indices and use $\Delta N$ and $\lambda$ for the observed number of events of a given node at a given inference time. The empirical distribution of $\lambda$ is represented by an ensemble. We also introduce a superscript $(s)$ for $s=1,\ldots,M$ to indicate the ensemble member where $M$ is the ensemble size. Suppose that the prior ensemble consists of $\lambda^{(s)}$ and has the ensemble mean $\bar{\lambda}$. Let $P$ be the ensemble variance and $P_r=P/\bar{\lambda}^2$ be the ensemble relative variance. If $\lambda^{(s)}$ is drawn from a gamma distribution with above mean and relative variance and $\Delta N$ has a Poisson likelihood function with mean $\lambda\delta t$, the posterior sample, denoted by $\lambda^{(s),a}$, is gamma-distributed with mean $\bar{\lambda^a}$ and relative variance $P_r^a$ that satisfy the following:
\begin{equation}\label{eq:updatemeanvariance}
\begin{split}
&\bar{\lambda^a} =\bar{\lambda} + \frac{\bar{\lambda}}{P_r^{-1}+\bar{\lambda}\delta t}(\Delta N-\bar{\lambda}\delta t)\\
&(P_r^a)^{-1} = P_r^{-1}+\Delta N.
\end{split}
\end{equation}
As shown in~\cite{Santitissadeekorn18}, the posterior sample that is consistent with~\eqref{eq:updatemeanvariance} satisfies the following formula:
\begin{equation}\label{eq:EnPGFmain}
\frac{\lambda^{(s),a}-\bar{\lambda^a}}{\bar{\lambda^a}}=\frac{\lambda^{(s)}-\bar{\lambda}}{\bar{\lambda}}+\frac{P_r}{P_r+(\Delta N)^{-1}}\biggl[\frac{\Delta N_e^{(s)}-\Delta\bar{N_e}}{\Delta\bar{N_e}}-\frac{\lambda^{(s)}-\bar{\lambda}}{\bar{\lambda}}\biggr],
\end{equation}
where $\Delta N_e^{(s)}$, called the perturbed observation, is independently drawn from a gamma distribution with mean $\Delta N$ and variance $1$ for $i=1\ldots,M$ and its sample mean is denoted by $\Delta\bar{N_e}$. Let $A^{(s),a}=\left(\frac{\lambda^{(s),a}-\bar{\lambda^a}}{\bar{\lambda^a}}\right)$, which is computed by the right-hand side of \eqref{eq:EnPGFmain}
Thus, we can (stochastically) update each ensemble member by 
\begin{equation}\label{eq:ensembleupdate}
\lambda^{(s),a}=\lambda^{(s)}+A^{(s),a}\bar{\lambda^a}.
\end{equation}
To clarify the algorithm below, we now recover all the node and time indices. We define a vector $\bq_k^i$ that collects all parameter estimate for the $i-$th node (i.e. $\mu^i, \beta^i, \alpha^{ij}$ for all $j$) at the time step $k$. The $s-$th ensemble member of the parameter vector is denoted by $\bq^{i,(s)}_k$ for $s=1,\ldots,M$ . Similarly, we define $\lambda^{i,(s)}_k$ denote the $s-$th ensemble member of the conditional intensity. We also define a vector $\Delta N_k:=\left[\Delta N^1_k,\ldots,\Delta N^m_k\right]$. The proposed algorithm can then be implemented for each node independently, which is concisely summarized below.
\begin{description}
  \item[Step 1] Given the ensemble from the $k-$th step $\left(\lambda^{i,(s)}_k,\bq^{i,(s)}_k\right)$ and $\Delta N_k$ we construct $\lambda^{i,(s),f}_{k+1}$ by substituting $\lambda^{i,(s)}_k$ and $\bq^{i,(s)}_k$ into~\eqref{eq:dh}.
  \item[Step 2] Once $\Delta N_{k+1}$ becomes available, we implement~\eqref{eq:EnPGFmain} to obtain $\lambda^{i,(s),a}_{k+1}$ as a result.
  \item[Step 3] The new ensemble of the conditional intensity $\lambda^{i,(s),a}_{k+1}$ is then used as the observation to update the model parameters of the $i-$th node. This is done by using Ensemble Kalman filter (EnKF), which will be explained below, and the result is a new ensemble of model parameters $\bq^{i,(s),a}_{k+1}$.
  \item[Step 4] Set $\lambda^{i,(s)}_{k+1}=\lambda^{i,(s),a}_{k+1}$ and $\bq^{i,(s)}_{k+1}=\bq^{i,(s),a}_{k+1}$. Repeat Steps 1-3 with the ensemble $\left(\lambda^{i,(s)}_{k+1},\bq^{i,(s)}_{k+1}\right)$ and $\Delta N_{k+1}$
\end{description}
\par
We now describe our implementation of Step 3 above, which is again separately done for each node at a given time. Thus, we temporarily suspend the node and time indices for a sake of convenience.  Let $\bar{\bq}_e$ be the ensemble mean of all $\bq^{(s)}$, i.e., $\bar{\bq}_e=M^{-1}\sum_{s=1}^M\bq^{(s)}$. We can compute the deviations $\bq^{(s)}-\bar{\bq}_e$ for each $s$ and stack them column-wise to form a scaled deviation matrix:
\begin{equation}\label{eq:deviation}
\bX=\frac{1}{\sqrt{M-1}}[\bq^{(1)}-\bar{\bq}_e,\ldots,\bq^{(M)}-\bar{\bq}_e]   
\end{equation}
Similarly, we can compute the matrix of scaled deviation for $\lambda^{(s)}$ and $\lambda^{(s),a}$, which will be denoted by $M-$dimensional row vectors $\bY$ and $\bY_a$, respectively. The sample $\bq^{(s)}$ can then be updated to obtain a new sample $\bq^{(s),a}$ similarly to the ``stochastic" EnKF formula~\cite{EnKFBook06}:
\begin{equation}\label{eq:KalmanEn_member}
\bq^{(s),a}=\bq^{(s)}+\bX\bY^\top(\bY\bY^\top+\bY_a\bY_a^\top)^{-1}(\lambda^{(s),a}-\lambda^{(s),f}).
\end{equation}
 Note that the inversion required by~\eqref{eq:KalmanEn_member} is trivial since it is a scalar. The pre-multiplication by $\bX$~\eqref{eq:KalmanEn_member} confines the new ensemble to the subspace spanned by the column of $\bX$. In other words, the algorithm lacks of necessary information to update any terms that has a non-zero projection onto the space orthogonal to the column space of $\bX$. This issue can be difficult to avoid in a large-dimensional problem, where the ensemble size tends to be much less than the dimension of the problem. Nevertheless, the above update gives the optimal solution in the subspace spanned by $\bX$, but in general not optimal in the parameter space. In the special case where the parameters are uncorrelated, the off-diagonal terms of $\bX$ can be reduced to zero and the innovation of intensity can be regressed on each parameter separately.

\section{Synthetic experiments}\label{s:results}
\subsection{Experiment 1: Perfect model}\label{sec:Perfect}
In this experiment, we demonstrate an application of EnPGF for the parameter estimation of~\eqref{eq:dh} with a particular focus on the network parameters $\alpha^{ij}$.
The data-generated model is the discrete-time Hawkes process~\eqref{eq:dh}; hence it is the same as the model used to propagate the ensemble as explained in the Section 3.
All parameters are assumed unknown except for $\delta t$. The parameter setting that imitates the covert network is of interest, i.e., some cell in the network has relatively low baseline rate but large $\alpha^{ij}$ for some $j\neq i$. To this end, we set the number of nodes $m$ to be 6, so there are 48 unknown parameters.

We set the true baseline rates to $\mu^j=2s_1$ for $j\in\{1,2,3,5,6\}$ and $\mu^4=0.75s_1$ for a constant $s_1>0$ and the decay rate to $\beta^j=5$ for all $j$. The excitation matrix $\alpha^{ij}$ is
\begin{equation*}
\alpha = s_2\begin{bmatrix}
1 &  0.5& 0.5& 0& 0& 0\\
1 &  1&  0.5&  0&  0& 0 \\
0&  1&  1/5&  0&  0& 0 \\
0&  1&  2.5&  0.5&  2.5& 0 \\
0&  0&  0& 0.4 &  1.5 &  0.5\\
0&  0&  0& 0.4 &  0.5 &  1.5.\
\end{bmatrix}
\end{equation*}
for a constant $s_2>0$. To investigate different scenarios, we consider 4 parameter sets: $s_1=s_2=1.5$, $s_1=1.5,s_2=0.5$, $s_1=0.5,s_2=1.5$ and $s_1=0.5,s_2=0.5$.
The scenario of $s_1=s_2=1.5$  will produce the highest rate of events. On the other hand, $s_1=0.5,s_2=0.5$ will have the lowest rate. The scenario $s_1=1.5,s_2=0.5$ will examine the ability of the filter to estimate the network structure when the excitation rate is low and the scenario $s_1=0.5,s_2=1.5$ will provide an investigation when the baseline is low, which may or may not affect the network reconstruction.
\par
The data is generated with the above parameters and $\delta t=0.1$ for 2000 steps. To implement EnPGF, an initial ensemble is required. We choose the ensemble size of $M=500$ and independently draw the baseline and decay rate for each node from a gamma distribution with mean of $4s_1$ and variance of 8. The parameter $\alpha^{ij}$ is drawn independently of a gamma distribution with mean of $s_2$ and variance of $1/4$.
 \par
\begin{figure}[htbp]
            \centerline{\includegraphics[width=\linewidth]{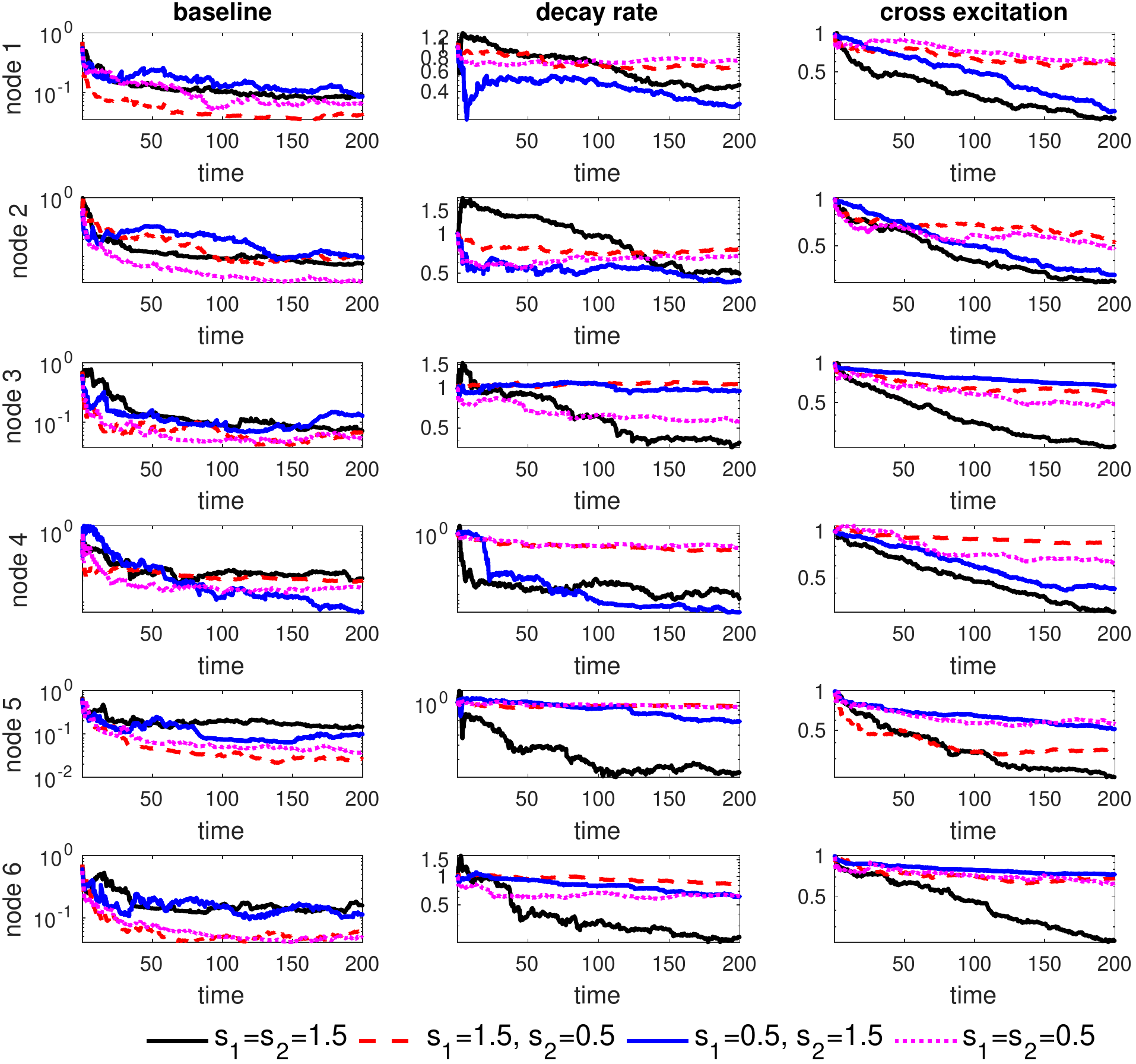}} \caption{The propagation of error reduction of all 4 cases, plotted in a logarithmic scale for the perfect model synthetic experiment. The value above $1$ indicates that the initial error is not reduced, while the value lower than 1 indicates that the degree of reduction of error.}\label{fig:toyerr}
        \end{figure}
\begin{figure}[htbp]
            \centerline{\includegraphics[width=\linewidth]{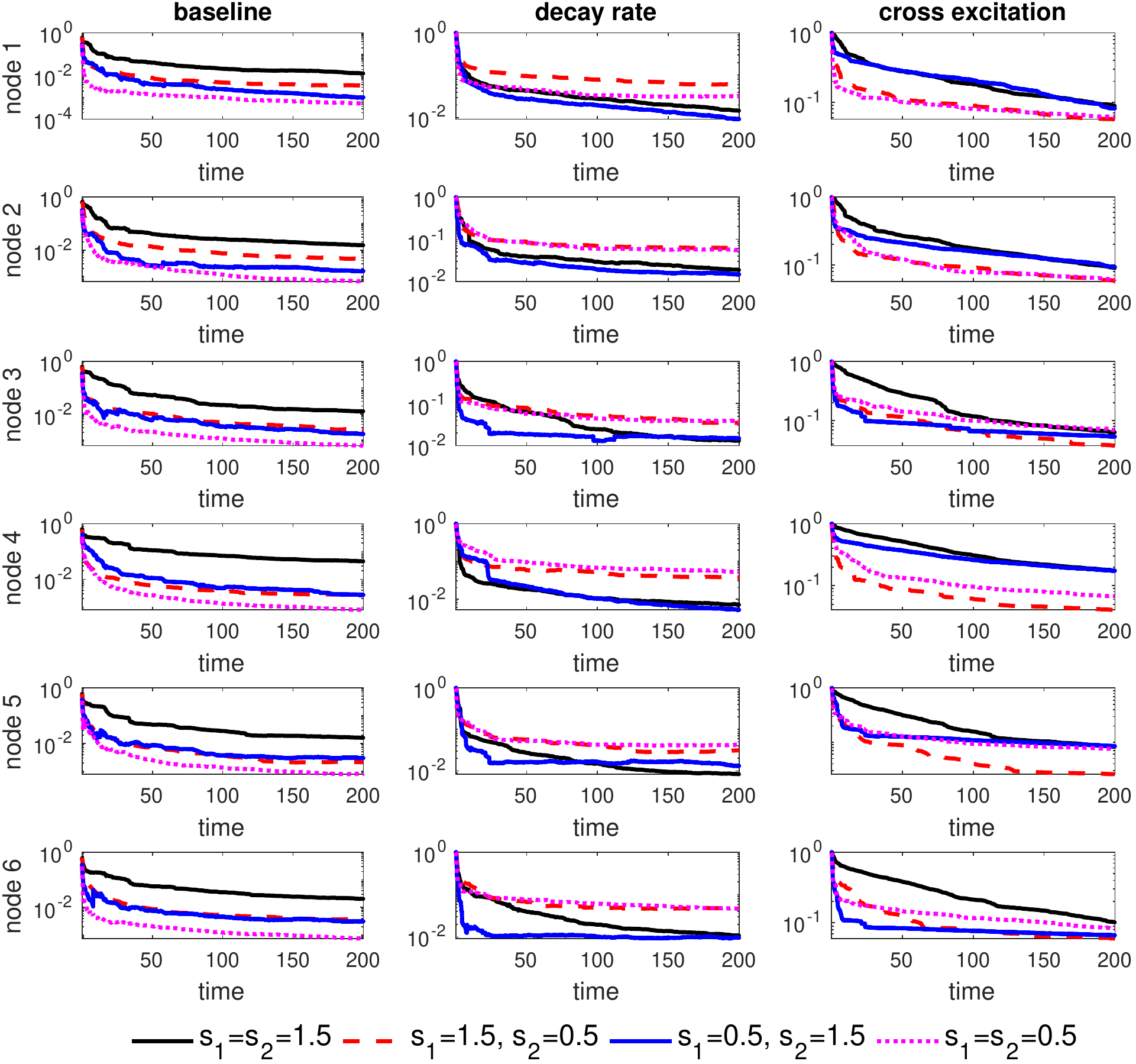}} \caption{The propagation of variance reduction of all 4 cases, plotted in a logarithmic scale for the perfect model synthetic experiment. The value above $1$ indicates that the initial variance is not reduced, while the value lower than 1 indicates that the degree of variance reduction.}\label{fig:toyvar}
        \end{figure}
\begin{figure}[htbp]
            \centerline{\includegraphics[width=\linewidth]{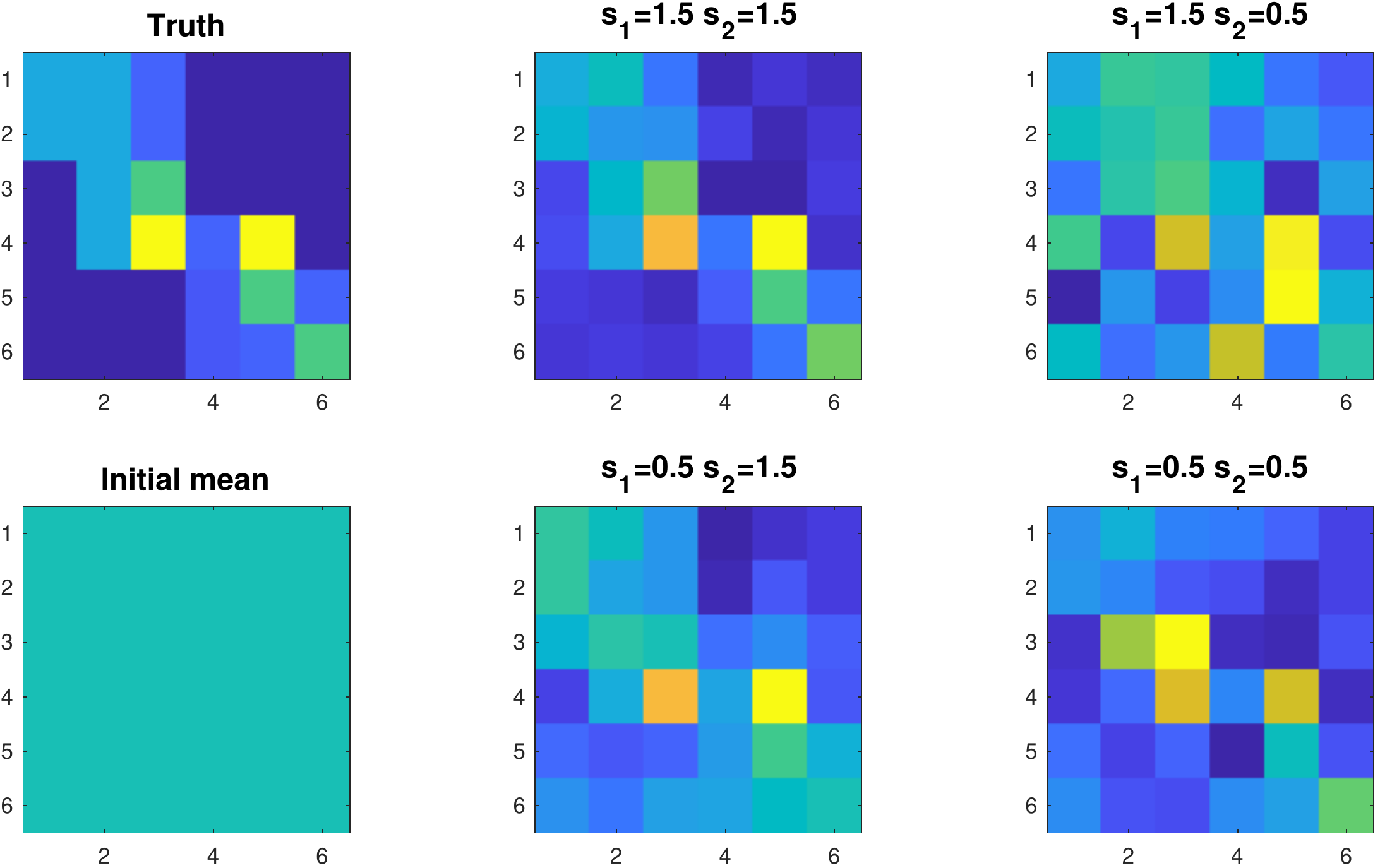}} \caption{Estimated network compared with the true network structure and initial condition for the perfect model synthetic experiment.}\label{fig:toynw}
        \end{figure}

The error reduction of the estimated parameters as the ensemble mean are shown in Figure~\ref{fig:toyerr}. More precisely, the root-mean-square error (rmse) of each parameter is computed and normalized by the rmse of the initial ensemble mean. Thus, the value of $1$ or above indicates no error reduction, whereas a value much smaller than $1$ indicates a significant error reduction. For each $i$-th node, the error of excitation rate is reported as the average error for $\alpha^{ij}$ for $j=1,\ldots,6$. Similarly, we present the variance reduction for each parameter in Figure~\ref{fig:toyvar}. The ensemble mean of the excitation matrix is also shown in Figure~\ref{fig:toynw}.
\par
As seen in~Figure~\ref{fig:toyerr}, a large error reduction is achieved for all nodes only for the scenario $s_1=s_2=1.5$. Nevertheless, the error reduction for the baseline rate is quite similar for all scenarios.  In terms of the variance reduction shown in Figure~\ref{fig:toyvar}, the reduction rate of the scenario $s_1=s_2=1.5$ is clearly slower than other cases despite its large error reduction. This is, however, to be expected since the Poisson-Gamma inference always reduce the variance in occasion of observing no events, but it can possibly increase the variance otherwise. Thus, the ensemble spread in terms of $\lambda^j$ is expected to be larger in the case of high baseline and excitation rates, which produces more non-zero observations than the other cases. Thus, even though the variance reduction is always obtained in the ensemble-based regression in~\eqref{eq:KalmanEn_member}, which uses the update of $\lambda^j$ as innovation term to estimate parameters, the variance reduction rate can be lower in the case of $s_1=s_2=1.5$. Figure~\ref{fig:toynw} shows the reconstructed network using the ensemble mean, and it is clear that the reconstruction is reliable when both baseline rate and excitation rate are large enough. Nevertheless, the strong values of $\alpha^{ij}$ can still be captured in all scenarios.
\par
\begin{figure}[htbp]
            \centerline{\includegraphics[width=\linewidth]{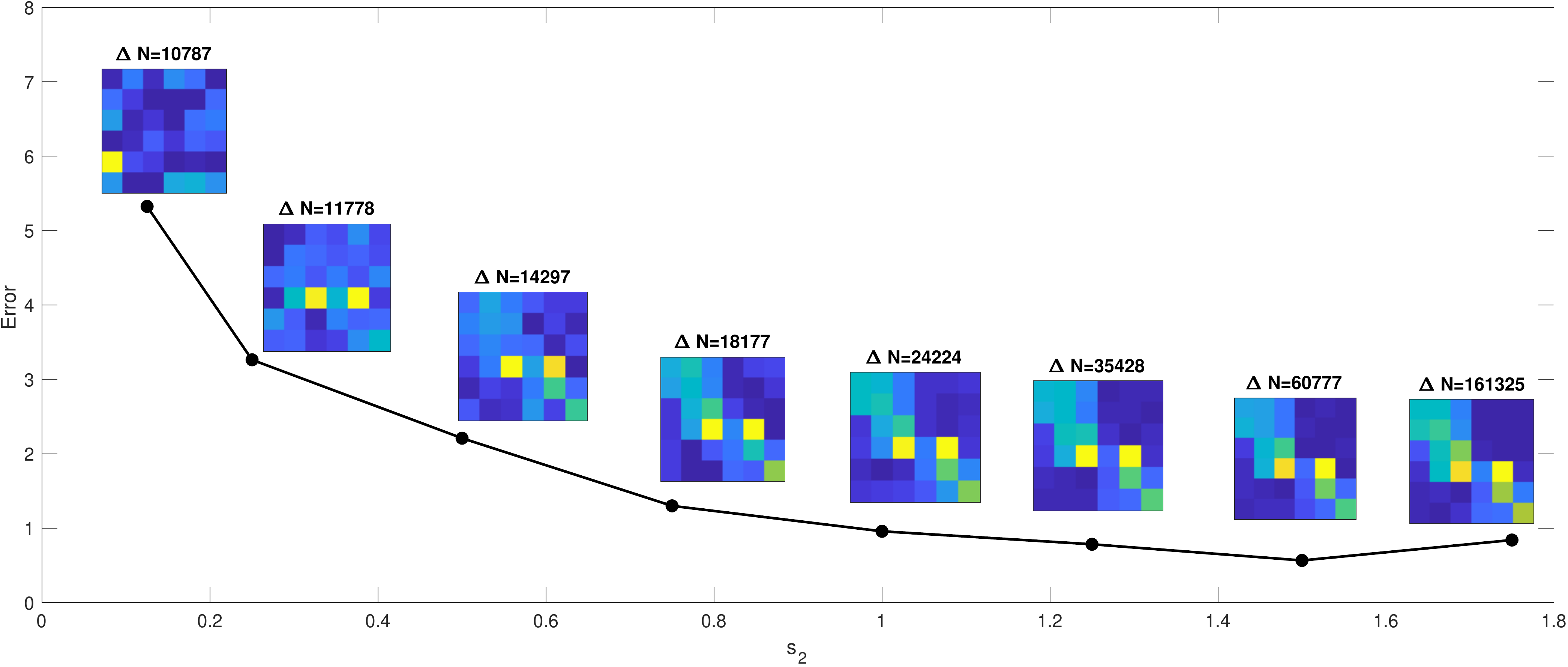}} \caption{Error of the excitation matrix measured by a scaled Frobenius norm for various values of $s_2$ for the perfect model synthetic experiment. The number of count in the test data, $\Delta N$, is also reported.}\label{fig:toyvarys2}
        \end{figure}
\begin{figure}[htbp]
            \centerline{\includegraphics[width=\linewidth]{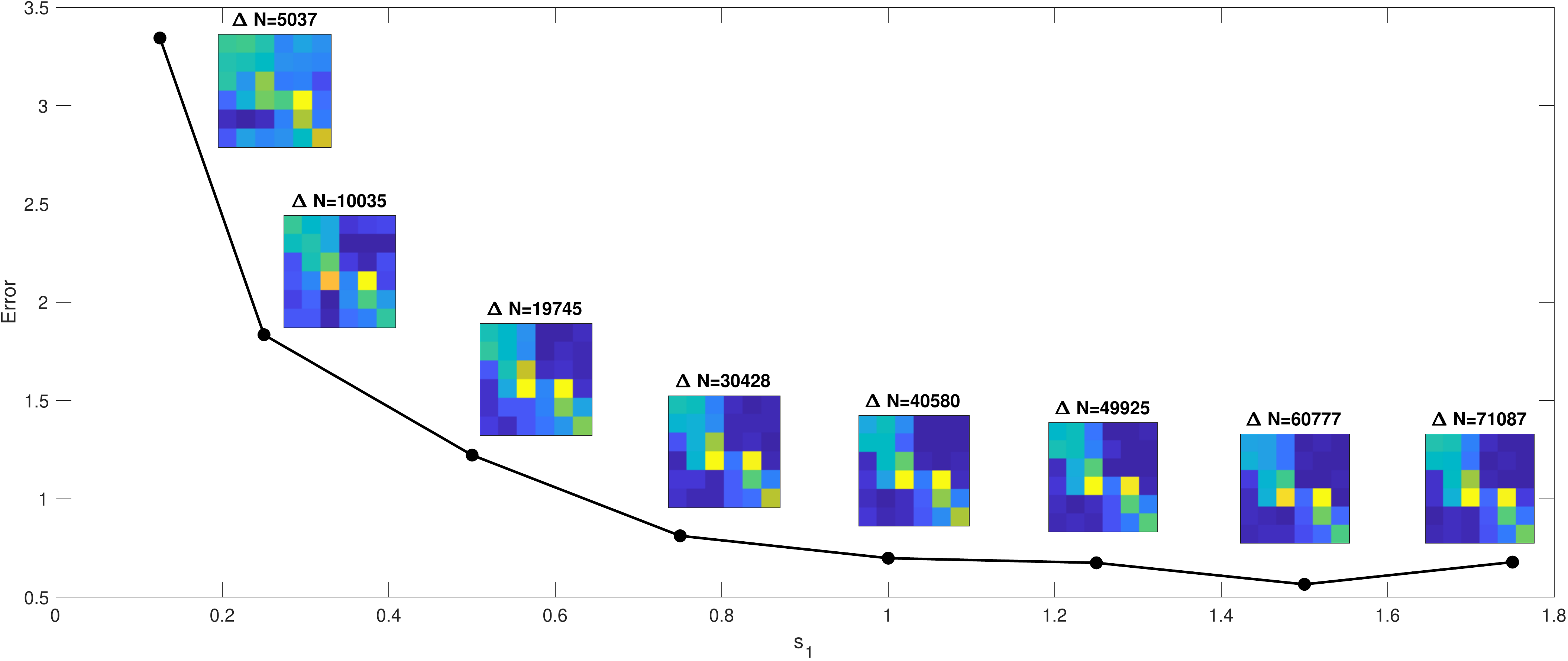}} \caption{Error of the excitation matrix measured by a scaled Frobenius norm for various values of $s_1$ for the perfect model synthetic experiment. The number of count in the test data, $\Delta N$, is also reported.}\label{fig:toyvarys1}
        \end{figure}

In Figure~\ref{fig:toyvarys2}, we focus on the impact of the excitation rate on the network reconstruction by fixing $s_1=1.5$ and varying $s_2$. The error of the excitation matrix is computed using the Frobenius norm, scaled by a factor of $1/s_2$. It is clear that the reconstruction deteriorates as the overall excitation rate of the network declines.
We also investigate the impact of baseline rate to the network reconstruction by fixing $s_2=1.5$ and varying $s_1$ as shown in Figure~\ref{fig:toyvarys1}. A deterioration trend is similarly observed, but the error is slightly smaller with slightly better tolerance to the lower value of baseline rate.

\begin{figure}[ht]
\begin{subfigure}{.5\textwidth}
  \centering
  \includegraphics[width=1\linewidth]{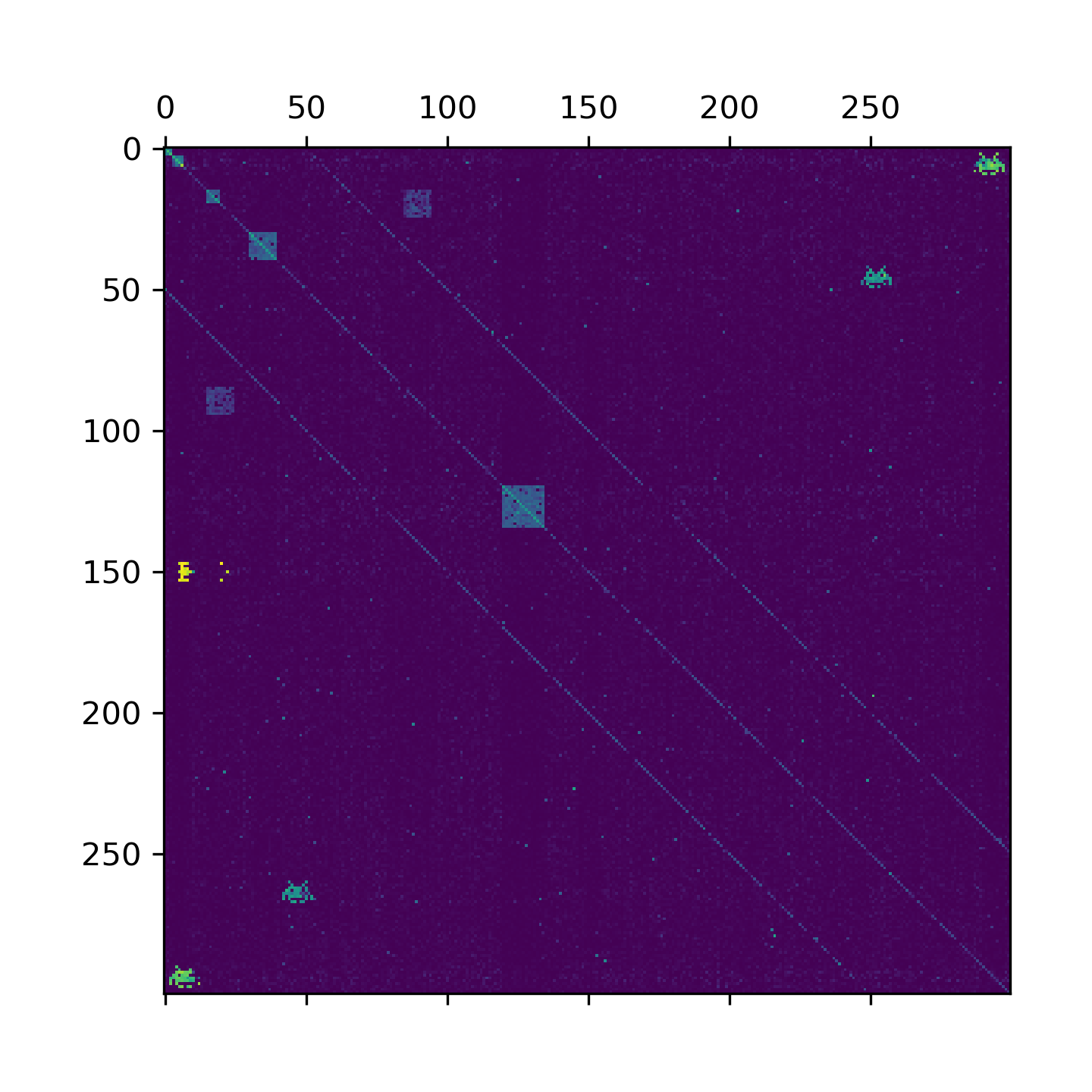}

  \label{fig:sub-first}
\end{subfigure}
\begin{subfigure}{.5\textwidth}
  \centering
  \includegraphics[width=1\linewidth]{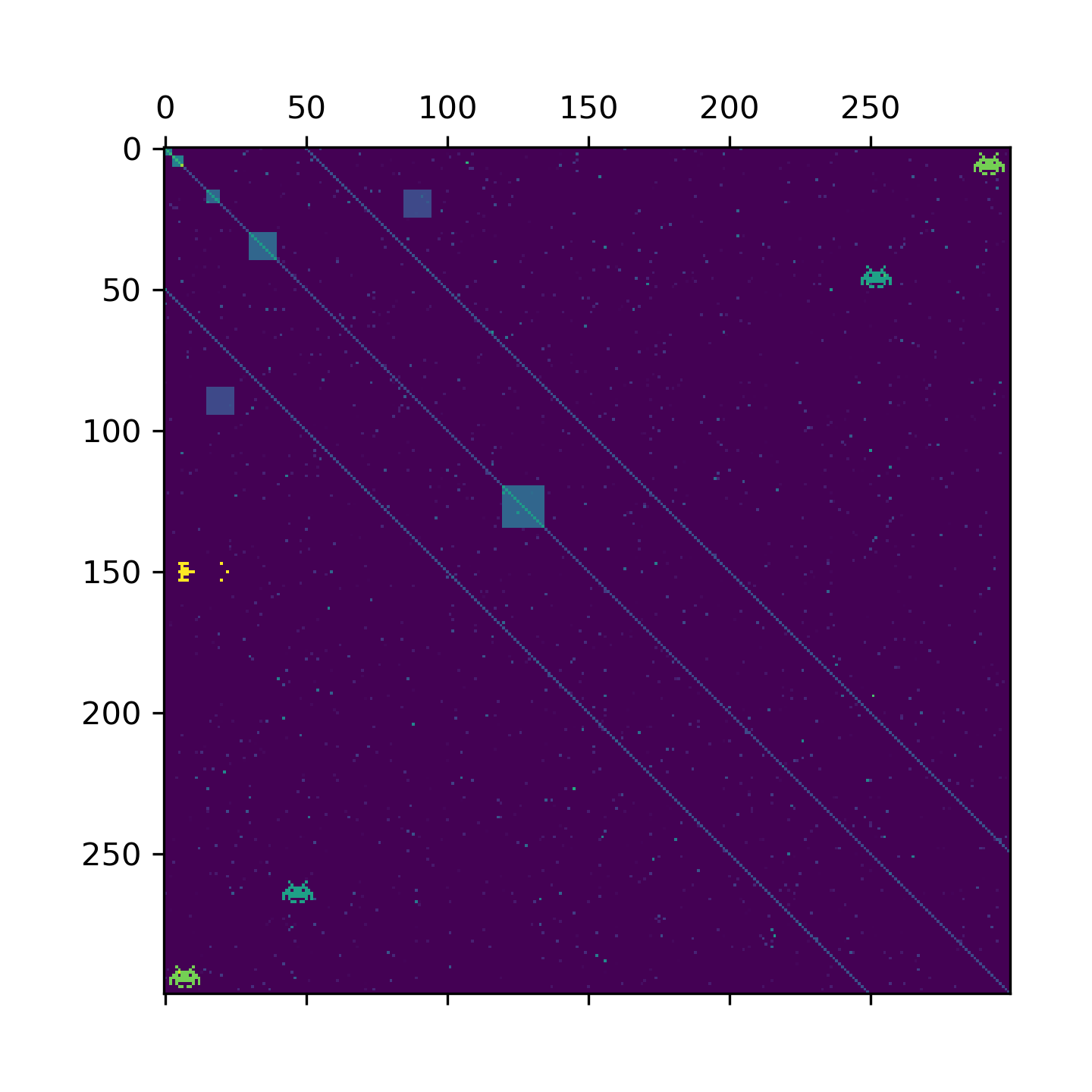}

  \label{fig:sub-second}
\end{subfigure}
\caption{(Left)Estimated excitation parameters for a network of 300 nodes. (Right)True parameters for a network of 300 nodes }
\label{fig:largesynthetic}
\end{figure}

We carry out a similar experiment for a large network of 300 nodes. A synthetic data generated from ``true" parameter values has a length of 150,000 steps with step size of $0.1$. All nodes have the true decay rate of $7$ and the baseline rate is drawn randomly from a gamma distribution with mean $20/3$ and variance $200/9$. The true excitation matrix is show in Figure~\ref{fig:largesynthetic} along with the network estimation, see Figure~\ref{fig:largesynthetic}.

\subsection{Experiment 2: Imperfect model}\label{sec:ABM}
In this experiment, we use an agent-based model (ABM) to generate the data but assuming~\eqref{eq:dh} for the prediction model; hence this is the imperfect model scenario. Therefore, the parameters used in the ABM cannot be taken as the ground truth for model parameters of~\eqref{eq:dh} in this experiment. We adopt the ABM developed in~\cite{short2008}, which allows for a diffusive effect of events across the spatial network through the excitation process and is a model for urban crime. In particular, an event at a node will increase the intensity of future events not only to itself but also its neighbours, which is fixed over time. The amount of excitation across the neighbouring nodes can be thought of a weighted, directed network. It is of interest to uncover this latent network by~\eqref{eq:dh}, so the main focus of this experiment is the estimation of parameter $\alpha^{ij}$.
\par
We briefly discuss the agent-based model (ABM) on a finite set of locations (or nodes) introduced in~\cite{short2008}. The ABM here consists of two main components: locations $s$ and agents who randomly generate an ``event" and move from one location to another. An agent decides to generate an event at a location $s$ between times $t$ and $t+\delta t$ according to
\begin{equation*}
p_s(t) = 1-\exp(-A_s(t))\delta t,
\end{equation*}
where $A_s(t)$ is called the ``attractiveness" of the location $s$ at time $t$.  This probability is assumed to have the following form:
\begin{equation*}
A_s(t+\delta t) = \mu_s+\biggl[(1-\eta)B_s(t)+\frac{\eta}{z}\sum_{s'\sim s}B_{s'}(t)\biggr](1-\omega\delta t)+\sum_{s'\sim s}w(s,s') E_s'(t).
\end{equation*}
The parameter $\mu_s$, $\omega$ and $\eta$ are static parameters, which are interpreted as the baseline, decay rate, diffusion rate, the number of nodes in the neighbour, excitation rate, respectively. The term $B_s(t)$ is a dynamic component with $B_s(0)=0$ and $E_s(t)$ is the number of events at the location $s$ between $t$ and $t+\delta t$. The parameter $w(s,s')$ defines the influence of the events at location $s'$ having on the location $s$ and $s'\sim s$ if $w(s,s')>0$. The parameter $z$ is the number of $w(s,s')>0$ for a given $s$. Therefore, the dynamic of attractiveness field is very similar to the multidimensional Hawkes process. The main difference is the absence of the diffusion in the Hawkes process. We define a matrix $\mathbf{W}$ with $w(s,s')$ on the row $s$ and column $s'$, which can be considered as an underlying excitation structure of this agent-based model.
\par
The movement probability of an agent from a location $s$ to $s''$ in the neighbour of $s$ is given by
\begin{equation*}
q(s,s'';t) =\frac{A_{s''}(t)}{\sum_{s'\sim s}A_{s'}(t)}.
\end{equation*}
If the agent who is at a location $s$ at time $t$ decides not to generate an event, the agent will move to one of the neighbouring locations of $s$ at time $t+\delta t$ according to the above probabilistic rule. Thus, the movement on the network is biased toward the location with a higher attractiveness. On the other hand, if the agent generates an event, the agent will be removed from the simulation. A new agent is also generated at each location at a rate $\Gamma$ (i.e. uniform distribution with mean $\Gamma$).
\par
\begin{figure}[htbp]
            \centerline{\includegraphics[scale=0.4]{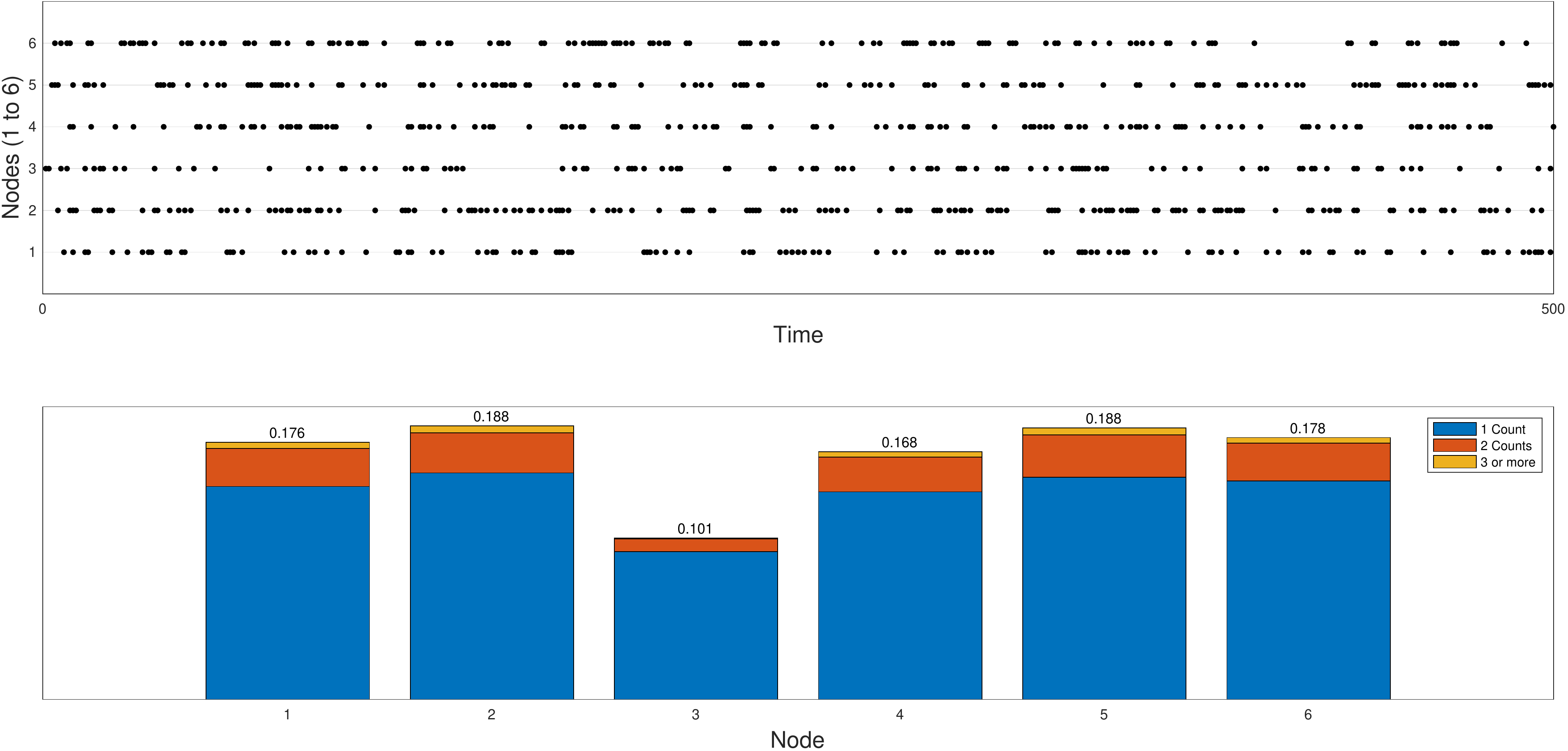}} \caption{Exploration of the ABM test data. (Top) The time-series of count data for the outgoing email only is shown. The dots represent the event of non-zero count. (Bottom) The height of the bar is proportional to the percentage of the outgoing email at each node, which is also reported at the tip of the bar. The height of the stacks in each bar is proportional to the events indicated in the plot (i.e. 1 count, 2 counts and 3 or more counts).}\label{fig:VisualABM}
        \end{figure}
\begin{figure}[htbp]
            \centerline{\includegraphics[scale=0.4]{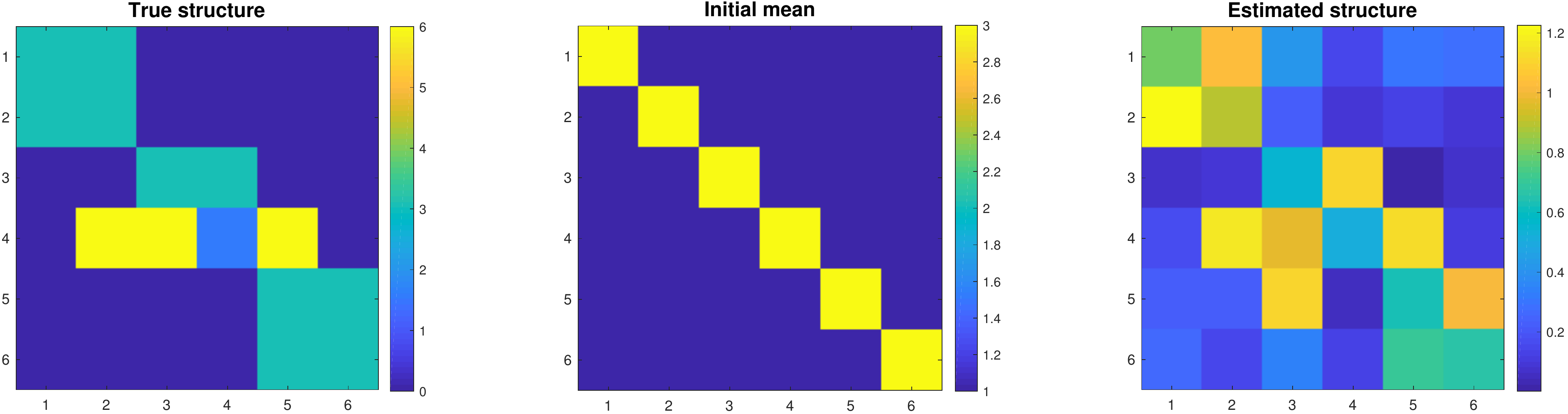}} \caption{Imperfect model experiment: (Left) ABM network structure. (Middle) Initial ensemble mean of the matrix $\alpha^{ij}$. (Right) Ensemble mean of $\alpha^{ij}$ at the final data assimilation step.}\label{fig:networkABM}
        \end{figure}

We test a problem with the network size of $m=6$  and use the following parameter values in our test experiment: $\omega=5, \eta=1/4, \delta t=1/10, \Gamma=3$.
The baseline rates $\mu_s$ for $s=1,\ldots,6$ are 2 for all locations except $\mu_4=1/2$. The ensemble size and initial ensemble are set up exactly the same as in the perfect model case, Section~\ref{sec:Perfect}. The excitation structure $\mathbf{W}$ is given below:
\begin{equation}
\mathbf{W}=
\begin{bmatrix}
3 & 3 & 0 & 0 & 0 & 0 \\
3 & 3 & 0 & 0 & 0 & 0 \\
0 & 0 & 3 & 3 & 0 & 0 \\
0 & 6 & 6 & 1.5 & 6 & 0 \\
0 & 0 & 0 & 0 & 3 & 3 \\
0 & 0 & 0 & 0 & 3 & 3 \\
\end{bmatrix}.
\end{equation}
The test data generated by the above parameter values for 39964 time steps is explored in Figure~\ref{fig:VisualABM}. Note that this network may be considered as a toy example of the covert network where the location $s=4$ has a relatively low frequency of the events, but it is the main influential location in the network. Although the excitation structure $\mathbf{W}$ in the ABM model cannot be taken precisely as a ground truth for the excitation parameters $\alpha^{ij}$ in~\eqref{eq:dh}, it is not unreasonable to expect a similar structure between them. As seen in Figure~\ref{fig:networkABM}, the estimated excitation network structures are similar despite the different mechanisms of the ABM and Hawkes model. The only significant misidentification in the structure is the excitation $w(5,3)$.

\section{Real-world data}\label{s:realworlddata}
\subsection{Ikenet data}
\begin{figure}[htbp]
            \centerline{\includegraphics[scale=0.5]{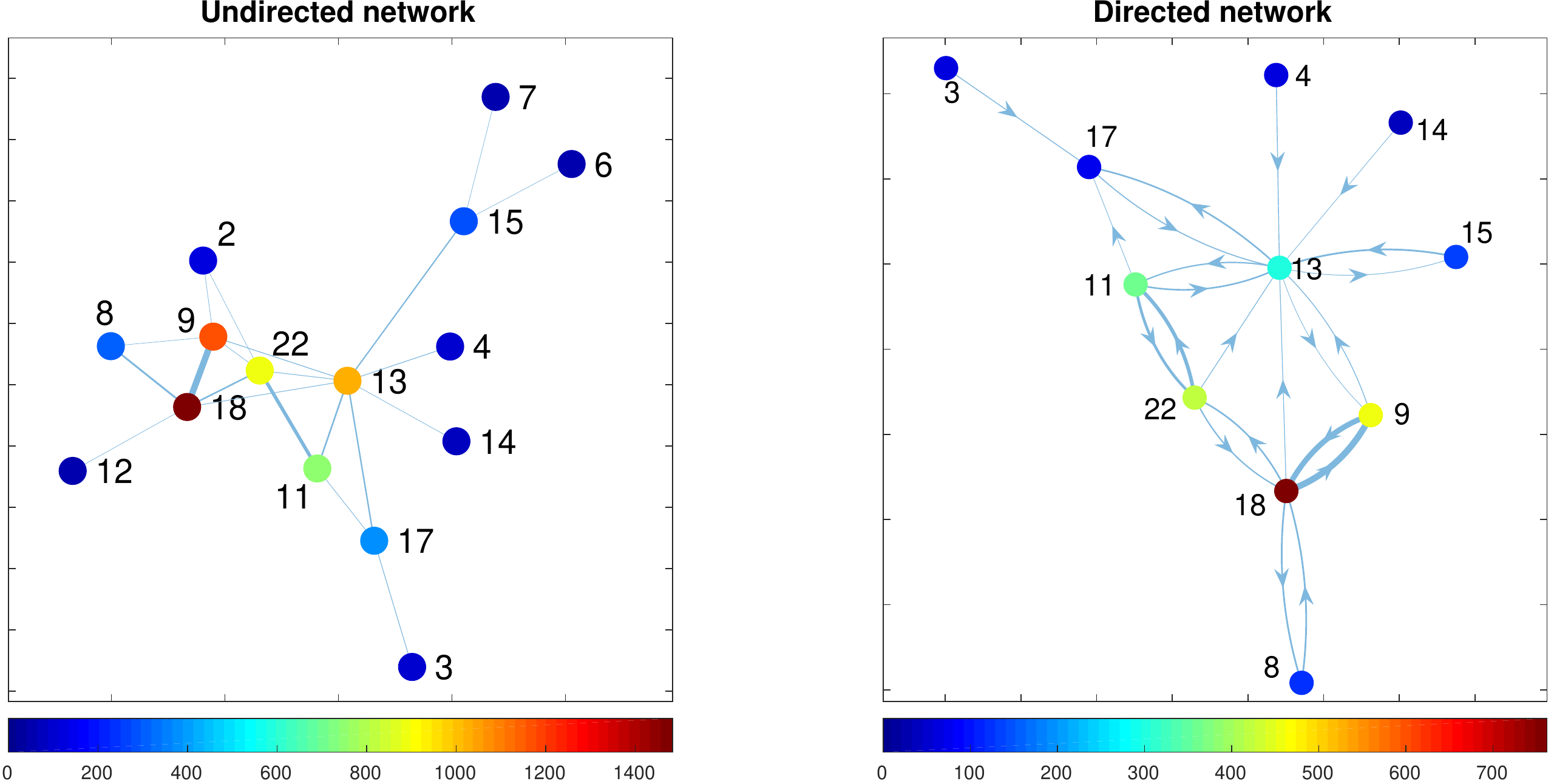}} \caption{(Left) Undirected subnetwork. The node colours vary with the degree of the node, and the edge size varies with the number of email communications between the corresponding pair of nodes. (Right) Directed subnetwork. The plot of network includes only edges with the weight larger than twice of the average weight.}\label{fig:IkenetG}
        \end{figure}

The Ikenet data set includes the timestamp of email exchanges between 22 anonymous volunteers (labelled by $\{1,\ldots,22\})$ over a one-year period beginning in May 2010.
There are 6678 email exchanges, not including those sending to oneself. The data can be visualized in a form of network where nodes represent the volunteers. In particular, we plot an undirected network whose links between any two volunteers are weighted by the total volume of email exchange between them and the directed network whose directional links are weighted by the number of email sent to a given volunteer. To aid a clear visualization, we plot only a subnetwork which includes only the edges that appear in the data more often than twice of the average number of appearances among all edges, see Figure \ref{fig:IkenetG}. It is clear that the pair $(9,18)$ is the most significant pair with similar weights in both directions.
\par
The Ikenet data was investigated in Fox et al.~\cite{Foxetal16} where the sending or receiving activity of each individual is modelled by the Hawkes process with a periodic baseline rate. The results show better Akaike Information Criteria (AIC) than a homogeneous Poisson model. The model parameters were fitted by Fox et al.~\cite{Foxetal16} using maximum likelihood estimation. In Zipkin et al.~\cite{Zipkin16}, the exchange pair $(i,j)$ for $i,j\in\{1,\ldots,22\}$ is modelled by a Hawkes process, in order to answer the missing data problem (the information of sender or receiver is lost in some of the existing edges). Their experiments suggest that the choice of the kernel function (exponential or power-law decay) makes an insignificant impact on the goodness-of-fit for the Ikenet data. Their work focused mainly on how the missing data would impact parameter estimation and how to fill in these missing data using the Hawkes process.
\par
\begin{figure}[htbp]
            \centerline{\includegraphics[scale=0.4]{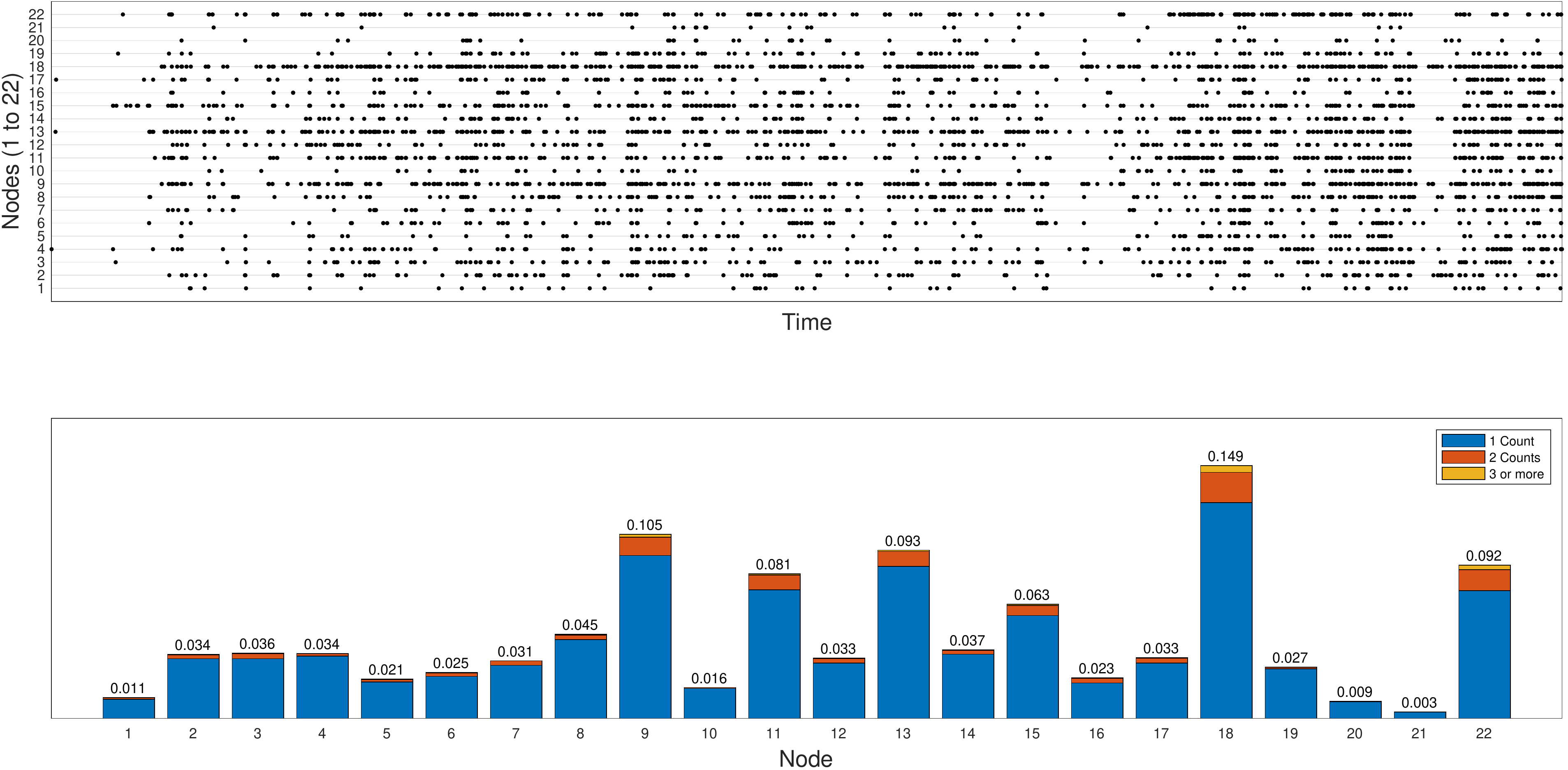}} \caption{Basic exploration of the IKEnet data. (Top) The time-series of count data for the outgoing email only is shown between time $0$ to 500 time steps. The dots represent the event of non-zero count. (Bottom) The height of the bar is proportional to the percentage of the outgoing email at each node, which is also reported at the tip of the bar. The height of the stacks in each bar is proportional to the events indicated in the plot (i.e. 1 count, 2 counts and 3 or more counts).}\label{fig:VisualIKE}
        \end{figure}

In our experiment, we will use only the sender data and convert the timestamp data into count data using a fixed time interval of $\delta t=0.1$ hour, resulting in 78832 steps in the data assimilation. The time-series of count data and the bar chart of the counts are shown in Figure~\ref{fig:VisualIKE}.

The goal is to estimate all parameters in~\eqref{eq:dh}. The ensemble of parameters has the ensemble size of $M=200$. The ensemble at in the final data assimilation step is used for a network inference. The result will show the most influential volunteers who make other people send emails more frequent than ``usual" (which is represented by the unknown baseline rate).

\par
\begin{figure}[htbp]
            \centerline{\includegraphics[width=\linewidth]{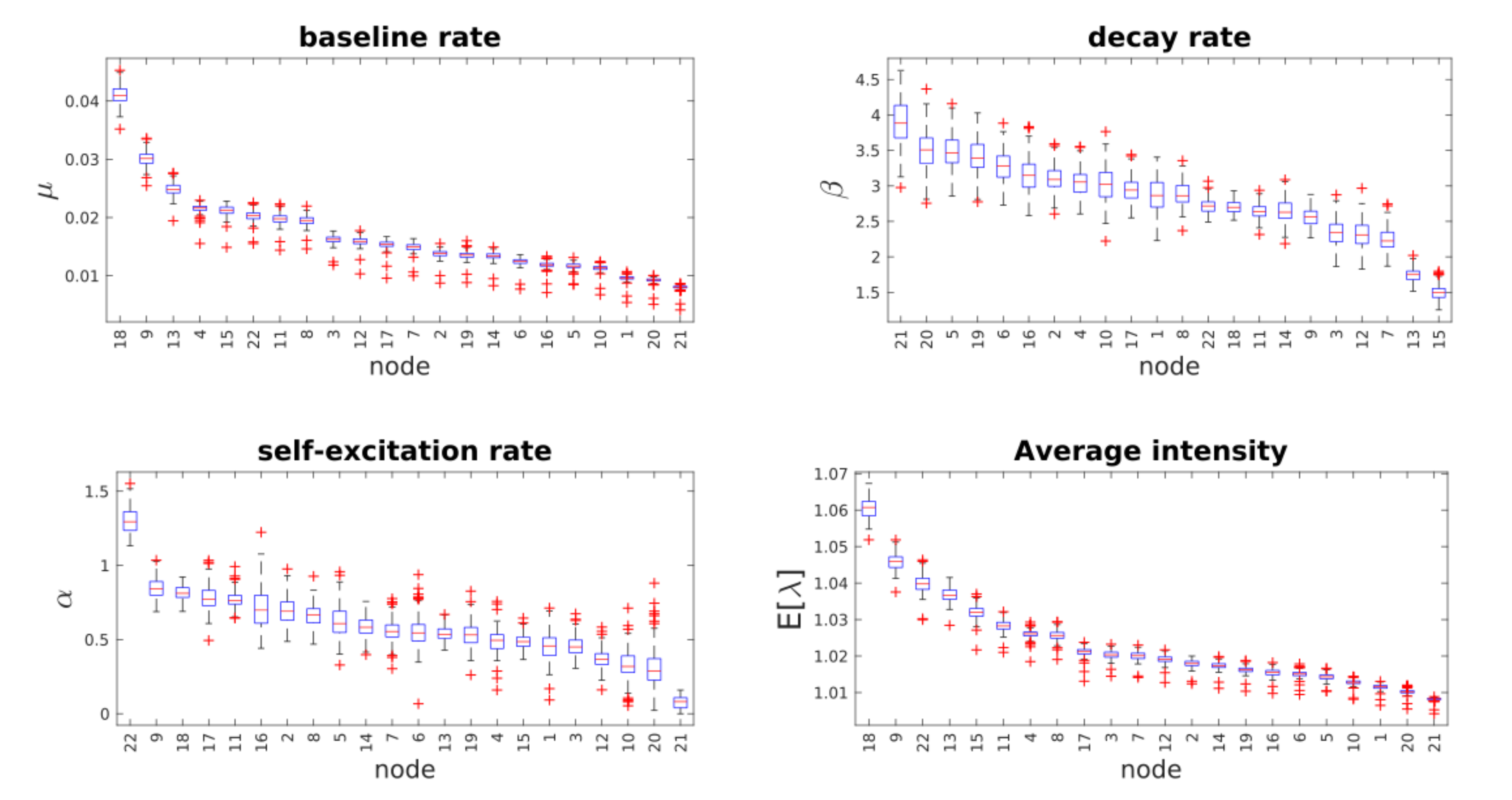}} \caption{The box plot for the ensemble of the baseline rate and decay rate.}\label{fig:Ikebox}
        \end{figure}
\begin{figure}[htbp]
            \centerline{\includegraphics[scale=0.75]{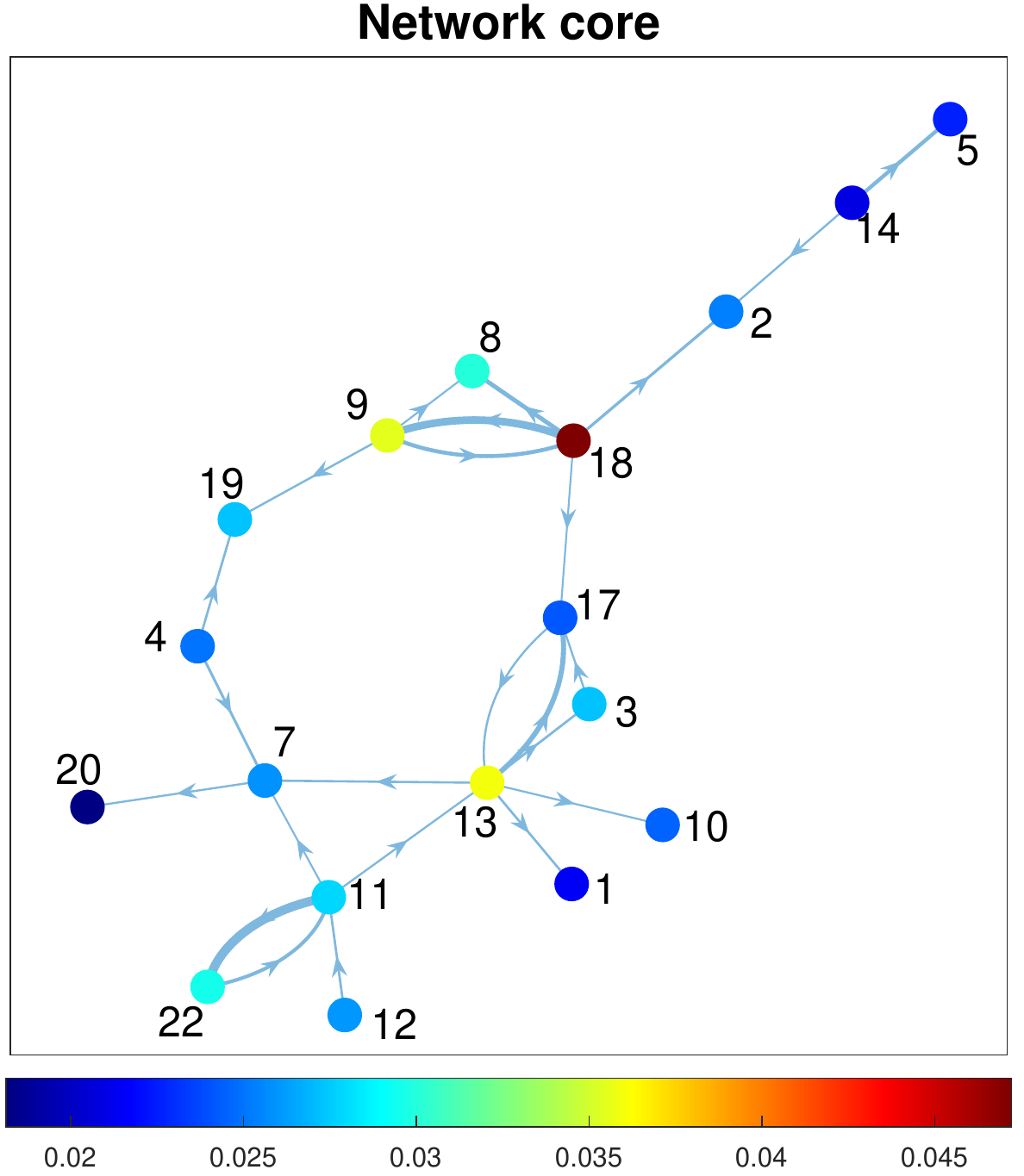}} \caption{The (mean) excitation network structure after removing some non-significant links and nodes. Each node is coloured according to the ensemble mean of the self-excitation rate.}\label{fig:IkeNW}
        \end{figure}

We now discuss the results obtained from the above setting. The boxplot, which summarizes the ensemble spread, of the baseline rate, decay rate and self-excitation rate (i.e. $\alpha_{ii}$ are shown in Figure~\ref{fig:Ikebox}. The nodes $18$ and $9$ have the highest baseline rates and average intensity rates, which agree with the visualization of the full data set, see Figure~\ref{fig:IkenetG}. However, the node with the highest self-excitation is the node 22. Again, our main interest is the influence network whose links are weighted by $\alpha_{ij}$ for $i\neq j$. To aid a clear visualization, we only plot the subnetwork here, see Figure~\ref{fig:IkeNW}. Some new insights can be gained from visualizing the influence network. For example, the node 18 has a significantly greater influence on the node 9 than the opposite direction despite the similar email exchange volumes between them, see again see Figure~\ref{fig:IkenetG}. A similar observation can also been seen from the pair of nodes 11 and 22. We emphasize that only the count time-series of the sender information is used here. Therefore, it is interesting to obtain the agreement of the influence network with the receiver information.

\par
\begin{figure}[htbp]
            \centerline{\includegraphics[scale=0.6]{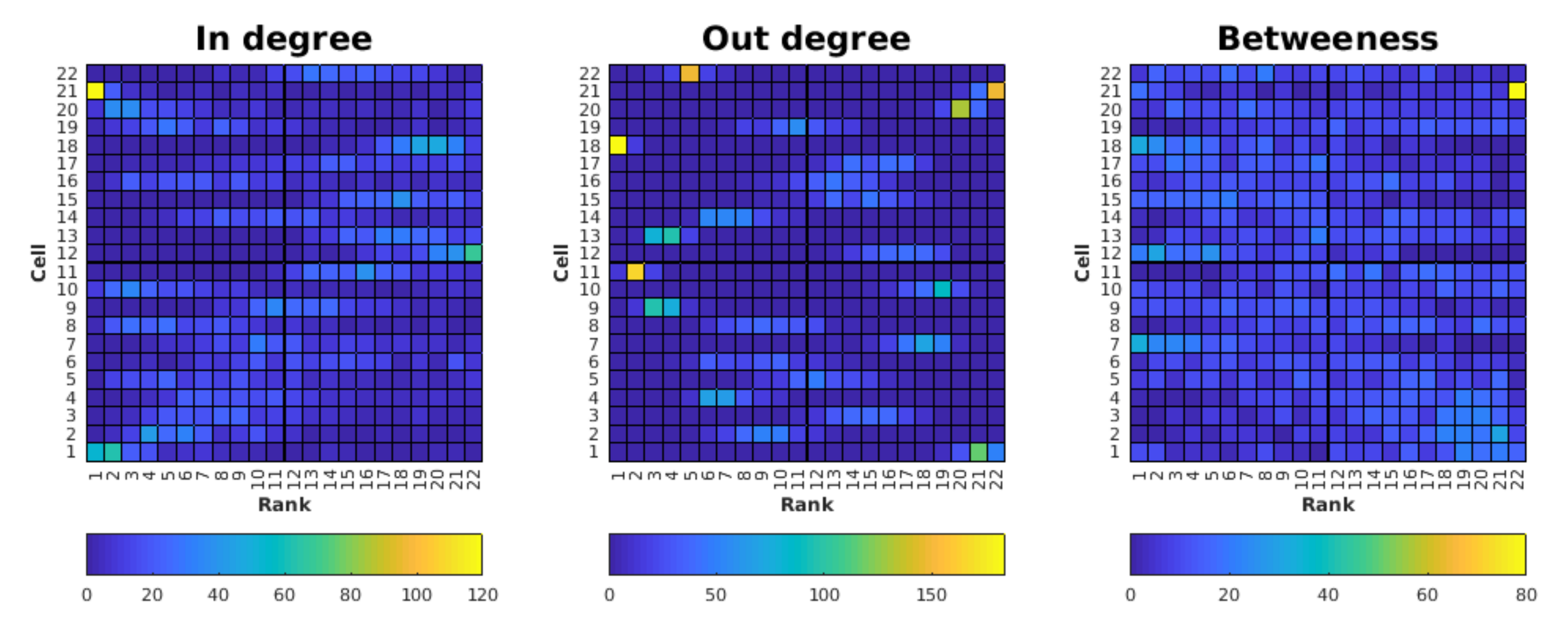}} \caption{The uncertainty of the centrality measures for the sender Network. The $(i,j)$ element of the colour map indicates the number of times (out of 200) the $i$-th rank is given to the $j$-th node.}\label{fig:Ikecentrality}
        \end{figure}

We can also explore the uncertainty of the network through the empirical distribution of the ensemble. We demonstrate this by investigating the uncertainty of various centrality measures of the network. We compute basic centrality measures of the influence network derived from each ensemble member, including out-degree, in-degree and betweenness centrality measures, as shown in Figure~\ref{fig:Ikecentrality}. In particular, we consider the proportion of the ensemble for which $i-$th node takes the $j-$th rank for a given centrality measure, i.e., the empirical distribution of the ranking. The result shows that some of the ranks can be quite certain, e.g., the top two ranks of the out-degree centrality taken by the nodes 11 and 18, respectively. The out-degree centrality may be used as a proxy of the most influential nodes in a sense of overall capacity to excite other nodes to sending out emails immediately after an email was sent out by the influential nodes. Thus, the node 18 has the most influence in the network. However, much remains uncertain for the in-degree centrality. The node 21, nonetheless, shows a high tendency of a large in-degree centrality but low out-degree centrality; hence it is the least influential node. The betweenness centrality for the sender network appears to be highly uncertain except for the last rank assigned mostly to the node 21.

\subsection{European email data}
\begin{figure}[htbp]
            \centerline{\includegraphics[width=\linewidth]{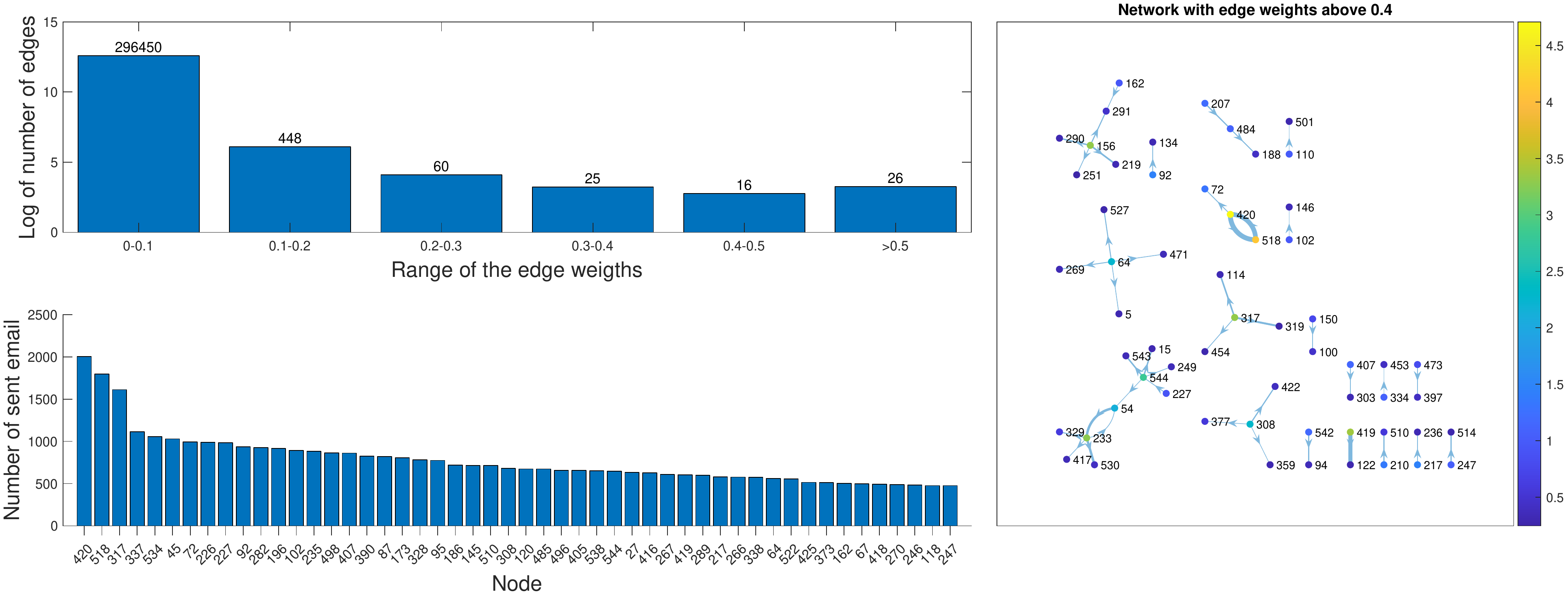}} \caption{(Upper left) The histogram of the edge weights found by the EnPGF algorithm, reported in a logarithm scale. Majority of edges is close to zero, indicating a sparse network. (Lower left)} The top 50 nodes that sent out the highest number of emails. (Right) The network ``core" that includes only edges with a weight above 0.4. The node colour corresponds to the total out degree of the node (i.e. the strength of influence in the current context). \label{fig:Euroemailresult}
        \end{figure}

We demonstrate the capability of EnPGF to deal with a relatively large network generated from a European email data. The European email data consists of time-series of email timestamps for 818 anonymous correspondents over 731 days. We ``clean-up" the data by removing days with almost no email activity (e.g. weekend, holidays, periods where data collection was disrupted) and the nodes with less than the total of 40 emails (receiving and sending combined) over 731 days.
This gives 514 nodes in total, and the timestamp data is aggregated into a unit of one hour for the test. Again, we consider only the outgoing email; hence, the recipients are unknown. We use an ensemble of 300 members in this analysis. Figure \ref{fig:Euroemailresult} shows the estimation result of the algorithm. It is clear that the discovered network is sparse in a sense that a large majority of the edges has insignificant (non-zero) weight. The network core, which include only edges with a weight above 0.4, clearly show a large mutual influence between node 420 and 518, which happen to be the two nodes that sent the highest number of emails. The estimation also detects a significant influence of the node 419 on node 122.

\section{Conclusion}\label{s:con}
We use an ensemble-based framework to infer the influence structure with the underlying multidimensional Hawke process models driven by count data. The proposed method shows promising results for both perfect model scenario and ABM-based model (i.e., the imperfect model). We also apply our method to the Ikenet email exchange data by using only the sending information. The results in Figure~\ref{fig:IkeNW} reveal interesting directional influence structure as interpreted by the Hawkes model~\eqref{eq:dh}. Some of the dominant influence links can be intuitively understood when comparing it with the full information in the data set (i.e. both sender and receiver information), see Figure~\ref{fig:IkenetG}. For instance, the strong influence of node 18 to node 9  is most likely attributed to the large number of email exchange between them. However, some influence links are more difficult to justify, e.g., why influence of node 11 on node 22 is much stronger than node 11 on node 17 despite similar numbers of email exchanges of the node pairs (11,22) and (11,17). The investigation of whether the influence network interpreted by the Hawkes model will fit into the real-world use is beyond the scope of our feasibility study here. We also highlight how the proposed method may be used for the uncertainty analysis of the network centrality measure. For this application context, the result helps to understand the uncertainty of how we might rank the influential nodes.
\par
The key limitations of the method used in this paper rest on the amount of data required to infer the influence network (we expect the amount of data required to grow exponentially with the size of the network) and whether influence is well modelled by the multivariate Hawkes process. These limitations will be true for any influence network inference that uses a multivariate Hawkes process. However, 
the proposed method has several advantages over existing methods. We are able to deal with count data and not just timestamp data that is common in datasets, we do not require knowledge of a physical network and due to the parallelization the method is able to carry out inference for large networks and data sets. Although our demonstration is very limited to the exponential-decay Hawkes process, our methodology offers a prospect of extension to several types of multidimensional Hawkes models. For example, extension of this work to reactive point processes that combine self-exciting effects from past events with self-inhibiting effects either from past events' history or from another event history would be fascinating. In the context of crime modelling, the crime intensity can be seen as self-excited by previous crime history and self-inhibited by police patrols (see \cite{Hefferan2016AdversarialPW} for adversarial patrolling); in the context of power grid failure, the failure intensity is self-excited by previous failure of the network and self-inhibited by inspections or maintenance (power grid maintenance \cite{ertekin2015}, water pipes maintenance \cite{8292799}). However, a strong nonlinearity between model parameters and the intensity dynamic of these Hawkes models in the general context may become an issue in the (ensemble-based) regression step. Additionally, while we assume the independency of parameters between the nodes,  it is possible to remove this restriction. This would though prohibit the parallelization of the algorithm; hence, reducing the feasibility of the method for a large-scale inference. Nevertheless, it might be possible to adapt the localization technique widely used for EnKF to improve computational efficiency as well as to avoid spurious correlation of the parameters. In the geophysical context, spurious correlations tend to be a long-range correlation between parameters of locations that are far apart.

\section*{Code availability}
Codes used to produce the results in this paper are available at

\url{https://github.com/naratips/EnPGF}

\section*{Data availability statement}
All data used to produce the results in this paper will be made available upon reasonable request.

\section*{Acknowledgment}
NS gratefully acknowledges the support of the UK Engineering and Physical Sciences Research Council for programme grant EP/P030882/1.

\section*{Appendix}\label{Appendix1}
We construct the update rule \eqref{eq:updatemeanvariance} based on the Poisson-Gamma conjugacy. Suppose that the prior distribution follows a gamma distribution with scale and shape parameters $a$ and $b$, respectively. Given an observation $\delta N$ drawn from a Poisson distribution with mean $\lambda\delta t$, the posterior distribution of $\lambda$ will have the scale and shape parameters $a+\delta N$ and $b+\delta t$. However, we depart from using the scale and shape parameter since we need to update the ensemble based on the ensemble mean and deviation from the mean. To this end, we will parameterize the gamma distribution with mean and relative variance instead. Recall that a gamma distribution with scale parameter $a$ and shape parameters $b$ has the mean $a/b$ and variance $a/b^2$. Thus, we can write
\begin{equation}
\begin{split}
E[\lambda\mid\Delta N] &= \frac{a+\delta N}{b+\delta t}=\frac{a}{b}+\frac{(a/b)}{a+(a/b)\delta t}\left(\Delta N-(a/b)\delta t\right)\\
&=E[\lambda]+\frac{E[\lambda]}{\frac{Var[\lambda]}{E[\lambda]^2}+E[\lambda]\delta t}\left(\Delta N-E[\lambda]\delta t\right).
\end{split}
\end{equation}
This yields the update equation of the mean in \eqref{eq:updatemeanvariance} by replacing $E[\lambda]$ by the ensemble mean and $Var[\lambda]$ by the ensemble variance.
The update rule of the relative variance in \eqref{eq:updatemeanvariance} can be similarly verified. 
As seen in \eqref{eq:ensembleupdate}, once we update the mean, we need to compute the deviation from the mean for all ensemble members to satisfy the update equation of the relative variance in \eqref{eq:updatemeanvariance}. We now show that we can stochastically update each ensemble member using the right-hand side of \eqref{eq:EnPGFmain} so that the updated ensemble will satisfy the relative variance in \eqref{eq:updatemeanvariance}.
\par
To verify \eqref{eq:EnPGFmain}, we first abbreviate each term in \eqref{eq:EnPGFmain} by
\begin{equation}
w^{(s)}:=\frac{\lambda^{(s),a}-\bar{\lambda^a}}{\bar{\lambda^a}},\quad u^{(s)}:=\frac{\lambda^{(s)}-\bar{\lambda}}{\bar{\lambda}},\quad t^{(s)}:=\frac{\Delta N_e^{(s)}-\Delta\bar{N_e}}{\Delta\bar{N_e}},\quad c:=\frac{P_r}{P_r+(\Delta N)^{-1}}.
\end{equation}
We make the following approximation:
\begin{equation}
\begin{split}
\frac{Var[\lambda^a\mid\Delta N]}{E[\lambda^a\mid\Delta N]^2}&\approx\frac{\frac{1}{M-1}\sum\limits_{s=1}^M(\lambda^{(s),a}-\bar{\lambda^a})^2}{\bar{\lambda^a}^2}=\frac{1}{M-1}\sum\limits_{s=1}^Mw^{(s)}=:P_r^a\\
\frac{Var[\lambda]}{E[\lambda]^2}&\approx\frac{\frac{1}{M-1}\sum\limits_{s=1}^M(\lambda^{(s)}-\bar{\lambda})^2}{\bar{\lambda}^2}=\frac{1}{M-1}\sum\limits_{s=1}^Mu^{(s)}=:P_r\\
\frac{Var[\Delta N_e\mid\Delta N]}{E[\Delta N_e\mid\Delta N]^2}&=\Delta N^{-1}\approx\frac{\frac{1}{M-1}\sum\limits_{s=1}^M(\Delta N_e^{(s)}-\bar{\Delta N_e})^2}{\bar{\Delta N_e}^2}=\frac{1}{M-1}\sum\limits_{s=1}^Mt^{(s)}.
\end{split}
\end{equation}
We may assume that given the observed value of $\Delta N$, the perturbed observation $\Delta N_e$ is conditionally independent to $\lambda$. Thus, based on the weak law of large number, we make the following approximation
\begin{equation}
\frac{1}{M}\sum\limits_{s=1}^Mu^{(s)}t^{(s)}\approx E[ut]=E[u]E[t]\approx\frac{1}{M}\sum\limits_{s=1}^Mu^{(s)}\frac{1}{M}\sum\limits_{s=1}^Mt^{(s)}=0.
\end{equation}
We square both sides of \eqref{eq:EnPGFmain} and then take the summation and divide by $M-1$. Applying all above approximations to the result yields
\begin{equation}\label{eq:check2}
\begin{split}
P_r^a&=E\left[\left(\frac{\lambda^{(s),a}-\bar{\lambda^a}}{\bar{\lambda^a}}\right)^2\right]=P_r-2\frac{P_r^2}{P_r+(\Delta N)^{-1}}+\frac{P_r^2}{P_r+(\Delta N)^{-1}},\\
&=P_r-\frac{P_r^2}{P_r+(\Delta N)^{-1}}.
\end{split}
\end{equation}

\bibliographystyle{plain}

\bibliography{EnPGFbib}

\end{document}